\documentclass[
  twocolumn,
  prl,
  showpacs,
  amsmath,
  amssymb,
  superscriptaddress,
  floatfix
]{revtex4}

\usepackage{bm}
\usepackage{graphicx}
\usepackage{color}

\newcommand{\be}{\begin{equation}}
\newcommand{\ee}{\end{equation}}
\newcommand{\beml}{\begin{subequations}}
\newcommand{\eml}{\end{subequations}}
\newcommand{\bea}{\begin{eqnarray}}
\newcommand{\eea}{\end{eqnarray}}
\newcommand{\ba}{\begin{array}}
\newcommand{\ea}{\end{array}}
\newcommand{\bpm}{\begin{pmatrix}}
\newcommand{\epm}{\end{pmatrix}}

\DeclareMathOperator{\Tr}{Tr}

\DeclareMathOperator{\sgn}{sgn}

\begin{document}

\title{Enhancement of superconductivity by Anderson localization}

\author{I.S.~Burmistrov}

\affiliation{ L.D. Landau Institute for Theoretical Physics, Kosygina
  street 2, 117940 Moscow, Russia} 

\author{I.V.~Gornyi}
\affiliation{
 Institut f\"ur Nanotechnologie, Karlsruhe Institute of Technology,
 76021 Karlsruhe, Germany
}
\affiliation{
 A.F.~Ioffe Physico-Technical Institute,
 194021 St.~Petersburg, Russia.
}
\affiliation{
 DFG Center for Functional Nanostructures,
 Karlsruhe Institute of Technology, 76128 Karlsruhe, Germany
}

\author{A.D.~Mirlin}
\affiliation{
 Institut f\"ur Nanotechnologie, Karlsruhe Institute of Technology,
 76021 Karlsruhe, Germany
}
\affiliation{
 DFG Center for Functional Nanostructures,
 Karlsruhe Institute of Technology, 76128 Karlsruhe, Germany
}
\affiliation{
 Inst. f\"ur Theorie der kondensierten Materie,
 Karlsruhe Institute of Technology, 76128 Karlsruhe, Germany
}
\affiliation{
 Petersburg Nuclear Physics Institute,
 188300 St.~Petersburg, Russia.
}

\begin{abstract}
Influence of disorder on the
temperature of superconducting transition 
($T_c$)
is studied within
the $\sigma$-model renormalization group framework. Electron-electron
interaction in 
particle-hole
and Cooper channels is taken into
account and assumed to be short-range. Two-dimensional
systems in the weak localization and antilocalization regime, as well
as systems near mobility edge are considered.
It is shown that in all these regimes the Anderson localization 
leads to strong enhancement of $T_c$ related to the
multifractal character of wave functions.  
\end{abstract}


\maketitle

Soon after the development of the microscopic theory of
superconductivity (SC) by Bardeen, Cooper, and Schrieffer (BCS) \cite{BCS},
the question of influence of disorder on SC
attracted a great deal of attention. It was
found \cite{AG59,Anderson59} that the diffusive motion of electrons
does not affect essentially the temperature $T_c$ of superconducting
transition, i.e the mean free path does not enter the expression for
$T_c$. This statement is conventionally called ``Anderson theorem''. 

Effects of disorder-induced Anderson localization
\cite{Anderson58} on SC were considered in
Refs.~\cite{MaLee85,KK85}. It was found
that, within the BCS approach, the SC in a disordered
system persists up to the localization threshold and even in the
localized regime near the Anderson transition. Furthermore,
Refs.~\cite{MaLee85,KK85} came to the conclusion that the mean-field
$T_c$ in these regimes remains unaffected by
disorder (i.e. the Anderson theorem holds).
In a parallel line of research, it was
discovered \cite{Maekawa82,Anderson83,Bulaevskii85} that an interplay of
long-range ($1/r$) Coulomb interaction and disorder leads to
suppression of $T_c$. These ideas were put on the solid basis by
Finkelstein \cite{Finkelstein87,FinkelsteinRev} who developed the $\sigma$-model
renormalization-group (RG) formalism.  

Recently, Feigelman {\it et al.} \cite{Feigelman07}, motivated by a
large body of experimental data~\cite{Exp}, returned to the problem of interplay
of disorder and SC. They 
found that 
the
eigenfunction 
multifractality near the localization
threshold strongly affects properties of a superconductor. Their
remarkable finding is that $T_c$ is dramatically enhanced: its
dependence on the coupling constant is no 
more exponential (as in the
conventional BCS solution) but rather of a power-law type. This result
was obtained on the basis of the BCS-type self-consistency equation,
with Cooper attraction being the only interaction included.  

In this paper we reconsider the problem of interplay of
SC and localization in the framework of the $\sigma$-model RG.
We take into account the interaction in all channels 
(singlet and triplet particle-hole, and Cooper). 
The key assumption that distinguishes our work from
Ref.~\cite{Finkelstein87} is the short-range character
of interaction that physically corresponds to a strong screening of
the long-range Coulomb interaction. We consider first two-dimensional
(2D) systems in the weak-localization (WL) and weak-antilocalization
(WAL) regimes
and find a strong enhancement of $T_c$ by disorder. We 
show that
this effect can be traced back to multifractality of wave functions.
We then extend the analysis to the vicinity of Anderson 
transitions in 2D and 3D (three-dimensional) systems. 

We begin by considering a 2D system of the orthogonal symmetry class
(i.e. with preserved spin-rotation symmetry). In the $\sigma$-model
formalism the problem is characterized by four running couplings: the
dimensionless resistance $t=2/\pi g$ (where $g$ is the conductivity of
the system
in
units of $e^2/h$) and three interaction constants $\Gamma_i$
corresponding to singlet 
particle-hole
($\Gamma_s$), triplet 
particle-hole ($\Gamma_t$), and
Cooper ($\Gamma_c$) channels \cite{Finkelstein87}. The RG treatment
yields a set of five coupled equations for these couplings and the
field renormalization constant $z$. It is convenient to switch from
$\Gamma_i$ to
normalized coupling constants $\gamma_i=\Gamma_i/z$ (typically, $\gamma_s, \gamma_c<0$ and $\gamma_t>0$); then the equation
for $z$ does not affect the remaining equations. The case of long-range Coulomb interaction corresponds to $\gamma_s=-1$.

Assuming weak short-range interaction,  
$|\gamma_i| \ll 1$, we have obtained the following RG system of equations (see
Supporting Material \cite{EPAPS}):
\bea
\label{e1}
{d\over dy} \left(\!\! \begin{array}{c} \gamma_s \\ \gamma_t \\ \gamma_c
\end{array}\!\! \right) &=& -{t\over 2} \left( \!\!\begin{array}{ccc} 
1 & 3 & 2 \\
1 & -1 & -2 \\
1 & -3 & 0\\
\end{array}\!\! \right) 
\left( \!\!\begin{array}{c} \gamma_s \\ \gamma_t \\ \gamma_c
\end{array}\!\! \right) - 
\left(\!\! \begin{array}{c} 0 \\ 0 \\ 2\gamma_c^2
\end{array}\!\! \right) \, ,  \\
dt / dy &=& t^2 \,     
\label{e2}\\
\label{e3}
d \ln z /dy &=& (t/2) (\gamma_s + 3\gamma_t +
2\gamma_c)\,. 
\eea
Here $y= \ln L$, where $L$ is the running RG length scale. We measure lengths in units of the microscopic scale 
where the RG \eqref{e1}-\eqref{e3} starts.
All the equations are written to the leading order in $t,\gamma_i\ll
1$. The first term on the r.h.s. of Eq.~(\ref{e1}) represents the
effects of disorder on interaction, while the second term is the
Cooper renormalization of $\gamma_c$. The r.h.s. of Eq.~(\ref{e2})
describes the WL effect.
We discard the
Altshuler-Aronov-type contribution in Eq.~(\ref{e2}) (describing renormalization
of disorder by interaction) since it is of higher order in
$\gamma_i$. 

Let us analyze the RG flow governed by Eqs.~(\ref{e1}) and
(\ref{e2}). Equation (\ref{e2}) decouples from the rest, yielding  
\be
\label{e4}
t^{-1}(y) = t_0^{-1} - y \,,
\ee
where the subscript $0$ refers to the bare value of the corresponding
coupling. This is the usual WL behavior. In the absence of
Eq.~(\ref{e1}) it would imply that strong Anderson insulator emerges
at the scale $y = t_0^{-1}$. We now turn to Eq.~(\ref{e1}). Let us
assume that the disorder is sufficiently strong compared to the
interaction, so that $t_0 \gg |\gamma_{i,0}|$. Then at the initial
portion of the RG flow we can neglect the Cooper renormalization term,
which leaves us with a linear system of equations. The corresponding
$3\times 3$ matrix has two eigenvalues: a positive one, $\lambda =
2t$ and a doubly degenerate negative one, $\lambda^\prime = -t$. Therefore,
as RG starts to operate, the vector formed by three couplings quickly
(at $y \sim 1$) approaches the eigenvector corresponding to
$\lambda$, i.e. $-\gamma_s = \gamma_t = \gamma_c \equiv \gamma$.
Projecting the system (\ref{e1}) onto this eigenvector, we get
\be
\label{e5}
d\gamma /dy = 2t \gamma - 2\gamma^2/3 \,.
\ee
We have checked that the neglected contributions do not affect the results in any essential way~\cite{EPAPS}.

\begin{figure}[t]
\centerline{\includegraphics[width=80mm]{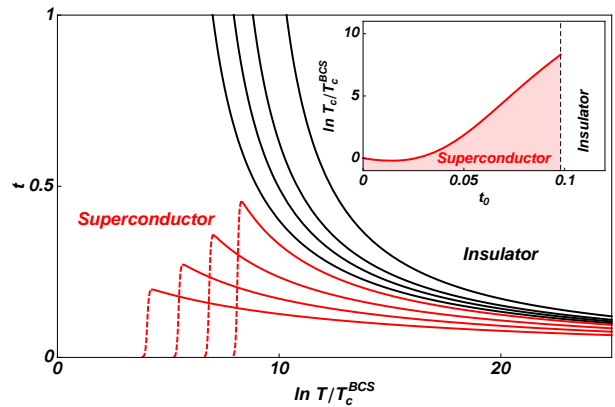}}
\caption{(Color online) Temperature dependence of resistivity $t$ near SIT in the 2D orthogonal symmetry class
from numerical solution of Eqs.~\eqref{e1} and \eqref{e2}
for $\gamma_{s0} = -0.005$, $\gamma_{t0}=0.005$, $\gamma_{c0}=-0.04$, and
$t_0 = 0.065, 0.075, 0.085, 0.095, 0.10, 0.105,0.11, 0.12$ (from bottom to 
top). 
The point where $\gamma_c$ diverges determines $T_c$ provided $t<1$.
Inset: dependence of $T_c$ on the bare resisitivity $t_0$.  
}
\label{Orthogonal_Figure}
\end{figure}

We will assume the initial value 
$\gamma_{0}= (-\gamma_{s,0} + 3\gamma_{t,0} + 2\gamma_{c,0})/6$ to
be negative which will imply SC (or at least a tendency
towards it)~\cite{FootnoteF}. This is, in particular, the case when the dominant bare
interaction is the Cooper attraction $\gamma_{c,0}<0$.   Solving 
then
Eq.~(\ref{e5}),
we find \cite{EPAPS} that
there are two distinct situations. If $|\gamma_{0}| \ll t_0^2$, the
resistance $t$ reaches a value of order unity when the interaction is
still weak. This means that when the scale is further increased
(i.e. the temperature is lowered), the system becomes an insulator. On
the other hand, if $|\gamma_{0}| \gg t_0^2$, the RG flow develops a
superconducting instability (blow up of the interaction) at a scale
where the resistance is still small, $t\ll 1$. This RG scale 
determines 
the temperature $T_c$ of superconducting transition,
\be
\label{e6}
T_c \sim \exp\{-2/t_0\} \, ,
\ee  
which 
is much 
higher
than the clean BCS value 
\be
\label{e7}
T_c^{BCS} \sim \exp\{- 1/{|\gamma_{c,0}|}\}
\ee 
in the considered regime of
sufficiently strong disorder,  $t_0 \gg |\gamma_{i,0}|$. When
the dimensionless resistance becomes smaller than the interaction,
Eq.~(\ref{e6}) crosses over into Eq. (\ref{e7}). 
We thus find that $T_c$ shows a non-monotonous dependence
on the disorder strength and gets strongly enhanced (by a
parametrically large factor in the exponential) in the intermediate
range of resistivities $t_0$, see Fig.~\ref{Orthogonal_Figure}. For given interaction strength, $T_c$ is the largest when the system approaches the superconductor-insulator transition (SIT) that takes place at $t_0^2 \sim |\gamma_{0}|$. 

It is important to emphasize a relation between this RG analysis of
the interacting problem and the multifractality of wave functions
in the non-interacting
theory. The wave function (or, equivalently, local density of states)
multifractality is a hallmark of criticality induced by 
Anderson localization \cite{Evers08}. It implies anomalous power-law
scaling of moments (and, more generally, of correlation functions) of wave
function amplitudes. In the field-theory language the corresponding
exponents are scaling dimensions of composite operators. In the fourth
order in wave function amplitudes $\psi$, which corresponds to the second
order with respect to the $\sigma$-model field $Q \sim
\psi\psi^\dagger $, there are two such operators. The dominant one has
a negative scaling dimension (i.e. it is RG relevant), $\Delta_2 <
0$, which is the most famous representative of the family of anomalous
dimensions $\Delta_q$ describing the wave function multifractality
spectrum \cite{Evers08,Wegner80,AKL}. The second operator has a positive dimension
$\mu_2>0$, which means that it is RG-irrelevant. It is worth
mentioning that, despite the RG-irrelevant character, $\mu_2$
controls the scaling behavior at Anderson transitions with short
range interaction \cite{irrelevant}. In terminology of Wegner
who pioneered the Anderson-localization multifractal
analysis \cite{Wegner80}, these exponents are denoted as $x_{2s}<0$ and $x_{2a}>0$,
respectively.  To the linear order in couplings $\gamma_i$, the RG equations of the
interacting theory should be controlled by scaling dimensions at the
non-interacting fixed point. We have verified that this is indeed the case.
Specifically, the exponent $\lambda = 2t$,
which is the positive eigenvalue of the matrix in (\ref{e1}) and which
shows up in Eq.~(\ref{e5}), is
nothing but the anomalous fractal dimension (with an opposite sign),
$\lambda=- \Delta_2$, while the second eigenvalue is the irrelevant
exponent, $\lambda^\prime = -t = - \mu_2$. Thus, the enhancement of $T_c$ is
intimately related to multifractality.  

\begin{figure}[t]
\centerline{\includegraphics[width=80mm]{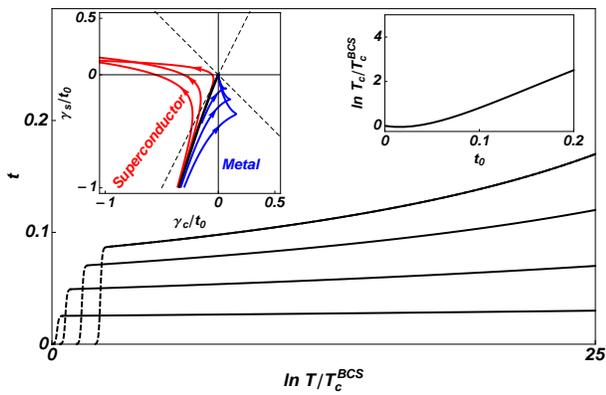}}
\caption{(Color online) Temperature dependence of resistivity $t$ in the 2D symplectic symmetry class
from numerical solution of Eqs.~\eqref{e8}-\eqref{e9} for $\gamma_{s0} = -0.005$, $\gamma_{c0}=-0.04$, and 
$t_0 = 0.02, 0.07, 0.12, 0.17$ (from 
bottom to 
top). 
Left inset: projection of RG flow~\eqref{e8}-\eqref{e9} on a plane $t= {\rm const}$. Separatrix (thick black curve) approaches the line $\gamma_s = 2\gamma_c$. Right inset: dependence of $T_c$ on the bare resisitivity $t_0$.
}
\label{SpinOrbit_Figure}
\end{figure}

Let us consider now a 2D system with strong spin-orbit
interaction. In this case the spin-rotation symmetry is broken and the
system belongs to the symplectic symmetry class. The change of the
symmetry leads to two important modifications of RG equations: (i) WAL
replaces WL, and (ii) triplet interaction channel gets suppressed
and can be discarded. The equations for the remaining interaction
constants $\gamma_s,\gamma_c$ and resistivity $t$ read  
\bea
\label{e8}
{d\over dy} \left( \begin{array}{c} \gamma_s \\  \gamma_c
\end{array} \right) &=&
 -{t\over 2} \left( \begin{array}{ccc} 
1 &  2 \\
1 & 0\\
\end{array} \right) 
\left( \begin{array}{c} \gamma_s \\ \gamma_c
\end{array} \right) - 
\left( \begin{array}{c} 0 \\ 2\gamma_c^2
\end{array} \right) \, ,
\\
dt/dy &=& -t^2/2 \,.
\label{e9}
\eea
The solution of Eq.~(\ref{e9}) describes the WAL flow 
\be
\label{e10}
t^{-1}(y) = t_0^{-1} + y/2 \,.
\ee
The eigenvalues of the linear part of the system (\ref{e8}) are again
related to the multifractality of wave functions (this time for the
symplectic symmetry class): $\lambda = t/2 = -\Delta_2$; $\lambda^\prime
= - t = - \mu_2$. The dominant 
eigenvalue $\lambda$ corresponds to the direction $-\gamma_s = \gamma_c \equiv
\gamma$,
see Fig.~\ref{SpinOrbit_Figure}.
Projecting the system onto this eigenvector, we get
\be
\label{e11}
d\gamma /dy = (t/2) \gamma - (4/3)\gamma^2 \,.
\ee
Assuming $\gamma_{0}=(-\gamma_{s,0}+2\gamma_{c,0})/3 <0$ and solving Eq.~(\ref{e11}),
we find the scale at which the coupling $|\gamma|$
becomes unity. 
This gives the transition temperature 
\be
\label{e12}
T_c \sim \exp\{-\mathcal{C}/(t_0|\gamma_{0}|)^{1/2}\} \,,
\ee
where a numerical prefactor $\mathcal{C}\sim 1$ depends on the ratio
$\gamma_{c,0}/\gamma_{s,0}$.  
Equation (\ref{e12}) is valid for $|\gamma_{0}| \ll t_0$; in the
opposite case the clean BCS result (\ref{e7}) is restored. The enhancement of $T_c$ in
Eq.~(\ref{e12}) 
is again due to multifractality represented by the eigenvalue $\lambda =
t/2 = -\Delta_2$ in Eq.~(\ref{e11}). The enhancement is less efficient
as compared to the orthogonal symmetry class, Eq.~(\ref{e6}), because
of antilocalizing behavior that leads to decrease of $t$ and thus to
weakening of multifractality (see Fig.~\ref{SpinOrbit_Figure}).

We consider now a system at the Anderson transition point.
This may be either a 2D or 3D symplectic class-system, 
or a 3D system of orthogonal symmetry. In all the
cases, after the initial (fast) part of the RG evolution, where the
Cooper term $\gamma_c^2$ is unimportant, 
$\gamma_s$ and (in the orthogonal case) $\gamma_t$ 
``adjust'' to $\gamma_c$ according to $\gamma_s = - \gamma_t = - \gamma_c$
(orthogonal) or  $\gamma_s = - \gamma_c$ (symplectic). So, the
main part of the RG evolution can be described by a single equation, in
analogy with Eqs.~(\ref {e5}) and (\ref{e11}):
\be
\label{e13}
d\gamma / dy = -\Delta_2 \gamma - a\gamma^2 \,, \qquad a\sim 1 \,,
\ee
where $\Delta_2 < 0$ is the fractal exponent for the given transition
point. 
In particular, $\Delta_2= - 1.7 \pm 0.05$ for the 3D
orthogonal-class and $\Delta_2= - 0.344\pm 0.004$ for the 2D symplectic-class Anderson transitions \cite{Evers08}.    
The SC will take place if $\gamma_{0} < 0$. 
Analyzing Eq.~(\ref{e13}), we find 
\be
\label{e14}
T_c^* \sim |\gamma_{0}|^{d/|\Delta_2|} \,.
\ee
for the transition temperature in $d$ spatial dimensions.
We see that at the Anderson transition point the enhancement of the
superconducting $T_c$
becomes even stronger than in
2D: $T_c$
is now a power-law (rather than
exponential) function of the interaction constant. 
Equation (\ref{e14}) agrees with the result of Ref.~\cite{Feigelman07}. 

\begin{figure}[t]
\centerline{\includegraphics[height=40mm]{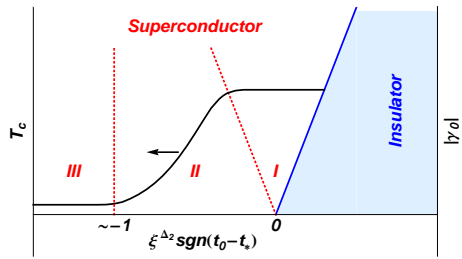}}
\caption{(Color online) Schematic phase diagram in the 
inter\-ac\-tion--dis\-order plane near the critical point. Solid (blue) line 
denotes SIT. Dependence of $T_c$ on the distance 
from 
$t_*$ at fixed value of $\gamma_{0}$ is shown by 
solid (black) curve. In the quantum critical regime (region I) $T_c$ 
is given by Eq.~\eqref{e14}. Away from criticality at $t\!<\! t_*$ (region 
III)  the BCS expression~\eqref{e7} holds. In the crossover regime 
(region II)  $T_c\! \sim\! \xi^{-3} \exp(-c_3 
\xi^{\Delta_2}/|\gamma_{c,0}|)$ in 3D and $T_c\! \sim\! \xi^{-2} 
\exp(-c_2 \sqrt{\xi^{\Delta_2}/|\gamma_{c,0}|})$
for the symplectic symmetry class in 2D \cite{EPAPS}.
}
\label{QC_Figure}
\end{figure}

The RG analysis allows us also to analyze the situation when the
system is slightly off the Anderson transition. It is convenient to
characterize the distance to the critical point $t_*$ by the correlation
(localization) length $\xi\! \sim\! |t_0-t_*|^{-\nu}$. The corresponding
energy scale is the one-particle level spacing in the correlation volume
$\delta_\xi\!\propto\! 1/\xi^{d}$. The result (\ref{e14}) retains its
validity in a vicinity of $t_*$
as long as $\delta_\xi
\lesssim T_c$. On the insulating side ($t_0\!>\! t_*$) the
condition $\delta_\xi\! \sim\! T_c$
determines the point of the superconductor-insulator quantum phase
transition (see Fig.~\ref{QC_Figure})~\cite{FootnoteFeigelman}. On the metallic side  ($t_0\!<\! t_*$), there is a crossover
regime extending from $\delta_\xi\!\sim\! T_c^*$ to $\delta_\xi \sim
1$ that provides a matching between the result (\ref{e14}) at the
Anderson-transition critical point and the clean BCS result \cite{EPAPS}.

We close the paper with several comments:

1) The true superconducting transition in 2D is of
Berezinskii-Kosterliz-Thouless (BKT) character, whereas we have
calculated the mean-field transition temperature $T_c$. It was found
however \cite{BKT} that the corresponding temperatures do not differ much,
$T_{\rm BKT}\simeq T_c$. Therefore, our result that shows an exponential
enhancement of $T_c$ is expected to hold for $T_{\rm BKT}$ as well.  

2) The key assumption of the above theory was the neglect of
long-range ($1/r$) Coulomb interaction. We can think of the following
situations when this should be justified: (i) a 3D material with a large
background dielectric constant $\epsilon$; (ii) a 2D system on a susbstrate (or
between two dielectrics) with large $\epsilon$; (iii) a 2D system with
a nearby screening metallic layer; (iv) a system of interacting
neutral fermions (i.e. cold atoms). 

3) A natural question is whether this effect has already been observed in
experiment. While one does see some
enhancement of SC by disorder in several materials, one
needs to argue that the Coulomb interaction is suppressed in order 
to attribute the increase of $T_c$ to the effect of localization. 
In particular, a non-monotonous dependence of $T_c$ on the
normal-state resistivity was found in
Refs.~\cite{Caviglia08,Kasahara09} in structures for which large
$\epsilon$ is expected. 

To summarize, we have developed a $\sigma$-model RG theory describing
interplay of SC and Anderson localization in a
disordered system with short-range interactions. This theory predicts
a strong enhancement of SC by Anderson 
localization in 
2D systems (at intermediate disorder) and near localization
transitions, implying a strongly non-monotonous dependence of $T_c$ on 
normal-state resistivity.
Remarkably, the localization physics 
responsible for increase of resistivity and thus driving the system towards an insulating state favors at the same time the SC.  It remains to be seen
whether this mechanism may be employed in practice to obtain structures with
strongly enhanced $T_c$. The key condition is a suppression of the
long-range component of the Coulomb interaction. 

We thank M. Feigelman, L. Ioffe, V. Kravtsov, B. Sacepe, and C. Strunk for useful discussions.
The work was supported by RFBR Grant 09-02-00247, Russian President Grant MK-296.2011.2, RAS Programs
``Quantum Physics of Condensed Matter'' and ``Fundamentals of nanotechnology and nanomaterials'', 
DFG-RFBR, and EUROHORCS/ESF.

\newpage

\begin{widetext}

\section{Supplementary Information}

\section{I.\, One-loop renormalization in the Finkel'stein NLSM}
\label{AppSecI}

\subsection{A.\, Nonlinear $\sigma$-model: Definitions}

The action of the Non-Linear Sigma Model (NLSM) is given as a sum of the non-interacting part, $S_\sigma$,
and contributions arising from the interaction in the particle-hole singlet, $S_{\rm int}^{(\rho)}$,
particle-hole triplet, $S_{\rm int}^{(\sigma)}$, and particle-particle (Cooper), $S_{\rm int}^{(c)}$, channels [S1,S2]:
\begin{gather}
S=S_\sigma + S_{\rm int}^{(\rho)}+S_{\rm int}^{(\sigma)}+S_{\rm int}^{(c)},
\end{gather}
where
\begin{gather}
S_\sigma = -\frac{g}{32} \int d\bm{r} \Tr (\nabla Q)^2 + 4\pi T z \int d\bm{r} \Tr \eta (Q-\Lambda) 
\label{Ss} \\
S_{\rm int}^{(\rho)}=-\frac{\pi T}{4} \Gamma_s \sum_{\alpha,n} \sum_{r=0,3}
\int d\bm{r} \Tr \Bigl [I_n^\alpha t_{r0} Q\Bigr ] \Tr \Bigl [I_{-n}^\alpha t_{r0} Q\Bigr ]
\label{Srho}\\
S_{\rm int}^{(\sigma)}=-\frac{\pi T}{4} \Gamma_t \sum_{\alpha,n} \sum_{r=0,3}\sum_{j=1}^3
\int d\bm{r} \Tr \Bigl [I_n^\alpha t_{rj} Q\Bigr ] \Tr \Bigl [I_{-n}^\alpha t_{rj} Q\Bigr ]
\label{Ssigma}\\
S_{\rm int}^{(c)}=-\frac{\pi T}{2} \Gamma_c \sum_{\alpha,n} \sum_{r=0,3}(-1)^r 
\int d\bm{r} \Tr \Bigl [I_n^\alpha t_{r0} Q I_{n}^\alpha t_{r0} Q\Bigr ]
\label{Sc}
\end{gather}
Here $g$ is the total Drude conductivity
(in units $e^2/h$ and including spin) and we use the following matrices
\begin{gather}
\Lambda_{nm}^{\alpha\beta} = \sgn n \delta_{nm} \delta^{\alpha\beta}t_{00}, \qquad
\eta_{nm}^{\alpha\beta}=n \delta_{nm}\delta^{\alpha\beta} t_{00}, \qquad
(I_k^\gamma)_{nm}^{\alpha\beta}=\delta_{n-m,k}\delta^{\alpha\beta}\delta^{\alpha\gamma} t_{00}
\end{gather}
with $\alpha,\beta$ standing for replica indices and $n,m$ corresponding to 
Matsubara fermionic energies $\epsilon_n = \pi T (2n+1)$.

The matrices 
\be
\label{trj}
t_{rj} = \tau_r\otimes s_j
\ee
operate in the
particle-hole (index $r$) and spin (index $j$) spaces,
with the corresponding Pauli matrices denoted by
\begin{gather}
\tau_0 = \begin{pmatrix}
1 & 0\\
0 & 1
\end{pmatrix},\qquad  \tau_1 = \begin{pmatrix}
0 & 1\\
1 & 0
\end{pmatrix}, \qquad \tau_2 = \begin{pmatrix}
0 & -i\\
i & 0
\end{pmatrix}, \qquad \tau_3 = \begin{pmatrix}
1 & 0\\
0 & -1
\end{pmatrix} , \\
s_0 = \begin{pmatrix}
1 & 0\\
0 & 1
\end{pmatrix},\qquad  s_1 = \begin{pmatrix}
0 & 1\\
1 & 0
\end{pmatrix}, \qquad s_2 = \begin{pmatrix}
0 & -i\\
i & 0
\end{pmatrix}, \qquad s_3 = \begin{pmatrix}
1 & 0\\
0 & -1
\end{pmatrix} .
\end{gather}
The matrix field $Q(\bm{r})$ (as well as the trace $\Tr$) acts in the replica, Matsubara, 
spin, and particle-hole spaces. It obeys the following constraints:
\begin{gather}
Q^2=1, \qquad \Tr Q = 0, \qquad Q^\dag = C^T Q^T C ,
\end{gather}
 where 
 \be
 C = i t_{12}, \qquad C^T = -C. 
 \ee
 It can be useful to represent $Q$ as $Q=T^{-1} \Lambda T$, with matrices $T$ obeying
 \begin{equation}
 C T^* = -T C,\qquad (T^{-1})^* C = -C T^{-1} . \label{TC}
 \end{equation}

In order to avoid notational confusion, it is instructive to compare our notation with that 
of the reviews by Finkel'stein~[S1]
and by Belitz and
Kirkpatrick~[S2].
First of all, these authors use different definitions of Pauli matrices:
\begin{align}
&\text{Ref. ~[S1]
:}\qquad\quad 
\tau^{F}_0 = \tau_0, \qquad \tau_j^{F} = i \tau_j,  \qquad\quad \sigma_0^{F} = s_0, \quad \sigma_j^{F} = s_j, \qquad\qquad j=1,2,3
\\
&\text{Ref. ~[S2]
:}\qquad\quad 
\tau^{BK}_0 = \tau_0, \quad \tau_j^{BK} = i \tau_j, \qquad s_0^{BK} = s_0, \quad s_j^{BK} = -i s_j, \qquad j=1,2,3
\end{align}
The interaction terms \eqref{Srho}, \eqref{Ssigma} and \eqref{Sc} coincide 
with terms in Eqs. (3.9a), (3.9b), and (3.9b) in Ref. [S1]
with
 the coupling constants
\begin{equation}
\Gamma_s = -\frac{\pi\nu}{4} Z, \qquad \Gamma_t=\frac{\pi\nu}{4}\Gamma_2, \qquad \Gamma_c =\frac{\pi\nu}{4} \Gamma_c,
\end{equation}
and with terms in 
Eqs. (3.92d), (3.92e), and (3.92f) in Ref.~[S2]
with
\begin{equation}
 \Gamma_s=K^{(1)}, \qquad  \Gamma_t=K^{(2)}, \qquad  \Gamma_c=K^{(3)}/2 ,
\end{equation}
where $\nu$ is the thermodynamic density of states including spin.
Finally, the parameters $g$ and $z$ in Eq.~(\ref{Ss}) are related by $g = 4\pi \nu D$ and $z= (\pi \nu/4) Z$ to the corresponding parameters introduced in Ref.~[S1]
and by $g=16/G$ and $z=H/2$ to those in Ref.~[S2].
Note that Ref.~[S1]
focuses on the case of unscreened (long-ranged) Coulomb interaction: hence the interaction
amplitude $\Gamma_s$ in the singlet particle-hole channel is expressed through the frequency renormalization factor $Z$ there.
In our case of a short-range interaction, these quantities are independent variables.

\subsection{B.\, Perturbation expansion}

We shall use the square-root parametrization
\begin{gather}
Q = W +\Lambda \sqrt{1-W^2}, \qquad W= \begin{pmatrix}
0 & w\\
\bar{w} & 0
\end{pmatrix} .
\end{gather}
We adopt the following notations: $W_{n_1n_2} = w_{n_1n_2}$ and $W_{n_2n_1} = \bar{w}_{n_2n_1}$ with $n_1>0$ and $n_2\leqslant 0$.
The blocks (in Matsubara space) obey
\begin{gather}
\bar{w} = -C w^T C,\qquad w = - C w^* C .
\end{gather}
The second equality implies that in the expansion $w^{\alpha\beta}_{n_1n_2}= \sum_{rj} (w^{\alpha\beta}_{n_1n_2})_{rj} t_{rj}$ some of the elements
$(w^{\alpha\beta}_{n_1n_2})$ are real and some are purely imaginary.

Expanding the action $S_\sigma$ to the second order in $W$, we find
\begin{gather}
S_\sigma^{(2)}=-\frac{g}{4} \sum_{rj} \sum_{n_1n_2}\sum_{\alpha\beta} 
\int \frac{d\bm{p}}{(2\pi)^d} \Bigl [p^2+\frac{32\pi Tz}{g}n_{12} \Bigr]
[w^{\alpha\beta}_{n_1n_2}]_{rj} [\bar{w}^{\beta\alpha}_{n_2n_1}]_{rj}
\end{gather}
Hence, the propagator becomes
\begin{gather}
\langle [w^{\alpha_1\beta_1}_{n_1n_2}]_{r_1j_1} [\bar{w}^{\beta_2\alpha_2}_{n_4n_3}]_{r_2j_2}\rangle = \frac{2}{g} D_p(n_{12}) \delta^{\alpha_1\alpha_2} \delta^{\beta_1\beta_2}\delta_{n_1n_3}\delta_{n_2n_4}\delta_{r_1r_2}\delta_{j_1j_2},
\\ 
D^{-1}_p(n_{12}) =p^2+\frac{32\pi Tz}{g}n_{12} . \label{Prop}
\end{gather}
Note that $\langle w w\rangle$ and $\langle \bar{w}\bar{w} \rangle$ are also non-zero. 
From Eq.~\eqref{Prop} we find
\begin{gather}
\langle \Tr a w \Tr b \bar{w} \rangle = \frac{32}{g} \sum_{n_1n_2}\sum_{\alpha\beta} \int \frac{d\bm{p}}{(2\pi)^d} D_p(n_{12}) \sum_{rj} [a_{n_2n1}^{\alpha\beta}]_{rj} [b_{n_1n2}^{\beta\alpha}]_{rj} \approx 4 Y \Tr \frac{1-\Lambda}{2} a \frac{1+\Lambda}{2} b, 
\\ 
Y =  \frac{2}{g} \int \frac{d\bm{p}}{(2\pi)^d} D_p(0) .
\end{gather}
In what follows we will use the following identity:
\begin{gather}
\langle \Tr A W \Tr B W \rangle = 2 Y \Tr \Bigl [ A B - \Lambda A \Lambda B - A C B^T C + \Lambda A \Lambda C B^T C\Bigr ] . \label{Aver}
\end{gather}

\subsection{C.\, Comments on interaction in the Cooper channel}

The Cooper channel interaction term $S_{\rm int}^{(c)} = (-\pi T \Gamma_c/2) O^{(c)}$ can be rewritten as
\begin{gather}
O^{(c)} = \frac{1}{2} \sum_{\alpha,n} \sum_{r=1,2} \sum_{j=0}^{3} \Tr \bigl [ t_{rj} L_n^\alpha Q \bigr ] \Tr \bigl [ t_{rj} L_n^\alpha Q \bigr ] , \qquad (L_n^\alpha)^{\beta \gamma}_{km} = \delta_{k+m,n}\delta^{\alpha\beta}\delta^{\alpha\gamma} t_{00}
\end{gather}
However, $\Tr \bigl [ t_{rj} L_n^\alpha Q \bigr ] =0$ for $j=1,2,3$ since
\begin{gather}
\Tr \bigl [ t_{rj} L_n^\alpha Q \bigr ] = \Tr \bigl [ t_{rj} L_n^\alpha Q \bigr ]^T = -\Tr \bigl [ C t^T_{rj} C L_n^\alpha Q \bigr ] =-\Tr \bigl [ C t^*_{rj} C L_n^\alpha Q \bigr ] = \begin{cases}
j=0, & \quad \Tr \bigl [ t_{r0} L_n^\alpha Q \bigr ] \\
j=1,2,3 & \quad -\Tr \bigl [ t_{rj} L_n^\alpha Q \bigr ] 
\end{cases} .
\end{gather}
Therefore, the operator $O^{(c)}$ describing the interaction in the Cooper channel is fully determined by the
Cooper-singlet channel:
\begin{gather}
O^{(c)} = \frac{1}{2} \sum_{\alpha,n} \sum_{r=1,2} \Tr \bigl [ t_{r0} L_n^\alpha Q \bigr ] \Tr \bigl [ t_{r0} L_n^\alpha Q \bigr ] . \label{Oc}
\end{gather}

\subsection{D.\, Background field renormalization}

\subsubsection{1.\, Singlet interaction in particle-hole channel: $S_{\rm int}^{(\rho)}=(-\pi T \Gamma_s/4) O^{(\rho)}$}

Performing transformation $Q \to T^{-1}_0 Q T_0$, we find with the help of Eq.~\eqref{Aver} and \eqref{TC}:
\begin{gather}
O^{(\rho)} = \sum_{\alpha,n}\sum_{r=0,3} \Tr \bigl [ I_n^\alpha t_{r0} Q \bigr ]  \Tr \bigl [ I_{-n}^\alpha t_{r0} Q \bigr ]
\to \left \langle  \sum_{\alpha,n}\sum_{r=0,3} \Tr \bigl [ I_n^\alpha t_{r0} T_0^{-1} Q T_0 \bigr ]  \Tr \bigl [ I_{-n}^\alpha t_{r0} T_0^{-1} Q T_0 \bigr ] \right \rangle \notag \\
=\left \langle  \sum_{\alpha,n}\sum_{r=0,3} \Tr \bigl [ T_0 I_n^\alpha t_{r0} T_0^{-1} W\bigr ]  \Tr \bigl [ T_0 I_{-n}^\alpha t_{r0} T_0^{-1} W \bigr ] \right \rangle =  2Y \sum_{\alpha,n}\sum_{r=0,3} \Tr \bigl [- I_n^\alpha t_{r0} Q_0 I_{-n}^\alpha t_{r0} Q_0 +I_n^\alpha t_{r0} Q_0 I_{n}^\alpha C t^*_{r0} C Q_0 \bigr ]  \notag \\
=  -2Y \sum_{\alpha,n}\sum_{r=0,3}\Bigl \{  \Tr \bigl [I_n^\alpha t_{r0} Q_0 I_{-n}^\alpha t_{r0} Q_0\bigr ] +(-1)^r \Tr \bigl [I_n^\alpha t_{r0} Q_0 I_{n}^\alpha t_{r0} Q_0 \bigr ]  \Bigr \} .
\end{gather}
Next, we represent the resulting expressions in terms of the operators entering in the interaction-induced part
of the action, \eqref{Srho}, \eqref{Ssigma}, and \eqref{Sc}:
\begin{gather}
\sum_{\alpha,n^\prime}\sum_{r=0,3}\Tr \bigl [I_{n^\prime}^\alpha t_{r0} Q_0 I_{-n^\prime}^\alpha t_{r0} Q_0\bigr ] = \frac{1}{2} \sum_{\alpha,n}\sum_{r=0,3}\sum_{j=0}^{3}\Tr \bigl [I_n^\alpha t_{rj} Q_0\bigr ] \Tr \bigl [ I_{-n}^\alpha t_{rj} Q_0\bigr ] = \frac{1}{2} O^{(\rho)} + \frac{1}{2} O^{(\sigma)}
\end{gather}
Finally, we see that the renormalized particle-hole singlet term is expressed through all three interaction 
operators
:
\begin{gather}
O^{(\rho)} \to O^{(\rho)} - Y O^{(\rho)} - Y O^{(\sigma)} -2 Y O^{(c)} . \label{ORho_R}
\end{gather}

\subsubsection{2.\, Triplet interaction in particle-hole channel: $S_{\rm int}^{(\sigma)}=(-\pi T \Gamma_t/4) O^{(\sigma)}$}

Performing transformation $Q \to T^{-1}_0 Q T_0$, we find with the help of Eq.~\eqref{Aver} and \eqref{TC}:
\begin{gather}
O^{(\sigma)} = \sum_{\alpha,n}\sum_{r=0,3}\sum_{j=1}^3 \Tr \bigl [ I_n^\alpha t_{rj} Q \bigr ]  \Tr \bigl [ I_{-n}^\alpha t_{rj} Q \bigr ]
\to \left \langle  \sum_{\alpha,n}\sum_{r=0,3} \sum_{j=1}^3\Tr \bigl [ I_n^\alpha t_{rj} T_0^{-1} Q T_0 \bigr ]  \Tr \bigl [ I_{-n}^\alpha t_{rj} T_0^{-1} Q T_0 \bigr ] \right \rangle \notag \\
=\left \langle  \sum_{\alpha,n}\sum_{r=0,3} \sum_{j=1}^3 \Tr \bigl [ T_0 I_n^\alpha t_{rj} T_0^{-1} W\bigr ]  \Tr \bigl [ T_0 I_{-n}^\alpha t_{rj} T_0^{-1} W \bigr ] \right \rangle 
\notag
\\
=  2Y \sum_{\alpha,n}\sum_{r=0,3} \sum_{j=1}^3 \Tr \bigl [- I_n^\alpha t_{rj} Q_0 I_{-n}^\alpha t_{rj} Q_0 +I_n^\alpha t_{rj} Q_0 I_{n}^\alpha C t^*_{rj} C Q_0 \bigr ]  \notag \\
=  -2Y \sum_{\alpha,n}\sum_{r=0,3} \sum_{j=1}^3\Bigl \{  \Tr \bigl [I_n^\alpha t_{rj} Q_0 I_{-n}^\alpha t_{rj} Q_0\bigr ] -(-1)^r \Tr \bigl [I_n^\alpha t_{rj} Q_0 I_{n}^\alpha t_{rj} Q_0 \bigr ]  \Bigr \} .
\end{gather}
Next,
\begin{gather}
\sum_{\alpha,n^\prime}\sum_{r=0,3}\sum_{j=1}^3 \Tr \bigl [I_{n^\prime}^\alpha t_{rj} Q_0 I_{-n^\prime}^\alpha t_{rj} Q_0\bigr ] = \frac{1}{2} \sum_{\alpha,n}\sum_{r=0,3}\Bigl \{3 \Tr \bigl [I_n^\alpha t_{r0} Q_0\bigr ] \Tr \bigl [ I_{-n}^\alpha t_{r0} Q_0\bigr ]
-\sum_{j=1}^{3}  \Tr \bigl [I_n^\alpha t_{rj} Q_0\bigr ] \Tr \bigl [ I_{-n}^\alpha t_{rj} Q_0\bigr ] \notag \\
= \frac{3}{2} O^{(\rho)} - \frac{1}{2} O^{(\sigma)} ,
\end{gather}
and
\begin{gather}
\sum_{\alpha,n^\prime}\sum_{r=0,3}(-1)^r\sum_{j=1}^3 \Tr \bigl [I_{n^\prime}^\alpha t_{rj} Q_0 I_{n^\prime}^\alpha t_{rj} Q_0\bigr ] = \frac{1}{2} \sum_{\alpha,n}\sum_{r=1,2}\Biggl \{3 \Tr \bigl [L_n^\alpha t_{r0} Q_0\bigr ] \Tr \bigl [ L_{n}^\alpha t_{r0} Q_0\bigr ]
-\sum_{j=1}^{3}  \Tr \bigl [L_n^\alpha t_{rj} Q_0\bigr ] \Tr \bigl [ L_{n}^\alpha t_{rj} Q_0\bigr ] \Biggr \}\notag \\
=\frac{3}{2} \sum_{\alpha,n}\sum_{r=1,2}\Tr \bigl [L_n^\alpha t_{r0} Q_0\bigr ] \Tr \bigl [ L_{n}^\alpha t_{r0} Q_0\bigr ]
 = 3 O^{(c)} .
\end{gather}
Finally, we find
\begin{gather}
O^{(\sigma)} \to O^{(\sigma)} - 3Y O^{(\rho)} + Y O^{(\sigma)} +6 Y O^{(c)} . \label{OSigma_R}
\end{gather}

\subsubsection{3.\, Interaction in Cooper channel: $S_{\rm int}^{(c)}=(-\pi T \Gamma_c/2) O^{(c)}$}

Performing transformation $Q \to T^{-1}_0 Q T_0$, we find with the help of Eq.~\eqref{Aver} and \eqref{TC}:
\begin{gather}
O^{(c)} = \sum_{\alpha,n}\sum_{r=0,3}(-1)^r \Tr \bigl [ I_n^\alpha t_{r0} Q I_{-n}^\alpha t_{r0} Q \bigr ] 
=\frac{1}{2} \sum_{\alpha,n}\sum_{r=1,2}\Tr \bigl [L_n^\alpha t_{r0} Q\bigr ] \Tr \bigl [ L_{n}^\alpha t_{r0} Q\bigr ]
\notag
\\
\to \frac{1}{2}\left \langle  \sum_{\alpha,n}\sum_{r=1,2} \Tr \bigl [ L_n^\alpha t_{r0} T_0^{-1} Q T_0 \bigr ]  \Tr \bigl [ L_{n}^\alpha t_{r0} T_0^{-1} Q T_0 \bigr ] \right \rangle \notag \\
=\frac{1}{2}\left \langle  \sum_{\alpha,n}\sum_{r=1,2}  \Tr \bigl [ T_0 L_n^\alpha t_{r0} T_0^{-1} W\bigr ]  \Tr \bigl [ T_0 L_{n}^\alpha t_{r0} T_0^{-1} W \bigr ] \right \rangle =  Y \sum_{\alpha,n}\sum_{r=1,2} \Tr \bigl [- L_n^\alpha t_{r0} Q_0 L_{n}^\alpha t_{r0} Q_0 +L_n^\alpha t_{r0} Q_0 L_{n}^\alpha C t^*_{r0} C Q_0 \bigr ]  \notag \\
=  -2Y \sum_{\alpha,n}\sum_{r=1,2} \Tr \bigl [L_n^\alpha t_{r0} Q_0 L_{n}^\alpha t_{r0} Q_0\bigr ] .
\end{gather}
Next,
\begin{gather}
\sum_{\alpha,n^\prime}\sum_{r=1,2} \Tr \bigl [L_{n^\prime}^\alpha t_{r0} Q_0 L_{n^\prime}^\alpha t_{r0} Q_0\bigr ] =
 \frac{1}{2} \sum_{\alpha,n}\sum_{r=0,3} (-1)^r\sum_{j=0}^{3}  \Tr \bigl [I_n^\alpha t_{rj} Q_0\bigr ] \Tr \bigl [ I_{n}^\alpha t_{rj} Q_0\bigr ]
\notag \\
= -\frac{1}{2} \sum_{\alpha,n}\sum_{r=0,3} (-1)^r\sum_{j=0}^{3}  \Tr \bigl [I_n^\alpha t_{rj} Q_0\bigr ] \Tr \bigl [ I_{-n}^\alpha C t^*_{rj} C Q_0\bigr ]
\notag \\
=-\frac{1}{2} \sum_{\alpha,n}\sum_{r=0,3} \Biggl \{\Tr \bigl [I_n^\alpha t_{r0} Q_0\bigr ] \Tr \bigl [ I_{-n}^\alpha t_{r0} Q_0\bigr ]
-
\sum_{j=1}^{3} \Tr \bigl [I_n^\alpha t_{rj} Q_0\bigr ] \Tr \bigl [ I_{-n}^\alpha t_{rj} Q_0\bigr ]\Biggr \}
= -\frac{1}{2} O^{(\rho)} + \frac{1}{2} O^{(\sigma)} . 
\end{gather}
Finally, we obtain
\begin{gather}
O^{(c)} \to O^{(c)} - Y O^{(\rho)} + Y O^{(\sigma)} . \label{OC_R}
\end{gather}
It is worthwhile to mention that only particle-hole singlet and triplet interaction operators are involved in the renormalized particle-particle term. It is due to the different properties of $I_n^\alpha$ and $L_n^\alpha$ matrices under the transposition operation: $(I_n^\alpha)^T=I_{-n}^\alpha$, $(L_n^\alpha)^T=L_n^\alpha$.

\subsection{E.\, One-loop renormalization group equations}

Using Eqs.~\eqref{ORho_R}, \eqref{OSigma_R} and \eqref{OC_R}, we find
\begin{gather}
S_{\rm int}^{(\rho)} + S_{\rm int}^{(\sigma)}+S_{\rm int}^{(c)} \to -\frac{\pi T}{4} \bigl (\Gamma_s-\Gamma_s Y -3 \Gamma_t  Y- 2\Gamma_c Y\bigr ) O^{(\rho)} \notag \\
- \frac{\pi T}{4} \bigl (\Gamma_t-\Gamma_s Y + \Gamma_t  Y+ 2\Gamma_c Y\bigr ) O^{(\sigma)} - \frac{\pi T}{2} \bigl (\Gamma_c - \Gamma_s Y+3 \Gamma_t  Y \bigr ) O^{(c)}
\end{gather}
Next, since
\begin{gather}
Y = \frac{2}{g} \int \frac{d\bm{p}}{(2\pi)^d} D_p(0) = \frac{1}{\pi g} \ln L/l
\end{gather}
we obtain the one-loop equations for the renormalization of the three interaction coupling constants 
\begin{align}
\frac{d\Gamma_s}{d\ln L} =& -\frac{1}{\pi g} (\Gamma_s+3\Gamma_t+2\Gamma_c), \\
\frac{d\Gamma_t}{d\ln L} =& -\frac{1}{\pi g} (\Gamma_s-\Gamma_t-2\Gamma_c), \\
\frac{d\Gamma_c}{d\ln L} =& -\frac{1}{\pi g} (\Gamma_s-3\Gamma_t).
\end{align}
The right-hand sides of these equations are linear in the interaction couplings, since we have neglected the
ladder resummation in the interaction propagators, as appropriate for weak couplings. 
The flow of the field renormalization constant $z$ follows from the renormalization 
of $\Gamma_s$  in view of the particle number conservation $d\Gamma_s/d\ln L = - dz/d\ln L$:
\be
\frac{dz}{d\ln L}=-\frac{d\Gamma_s}{d\ln L} = \frac{1}{\pi g} (\Gamma_s+3\Gamma_t+2\Gamma_c).
\ee
Introducing $\gamma_{s,t,c} = \Gamma_{s,t,c}/z$ and taking into account the definition of $t=2/\pi g$, 
we arrive at the linearized RG equations of the main text [see equations \eqref{e2.1}-\eqref{e2.5} below]:
\begin{align}
\frac{d\gamma_s}{d\ln L} =& -\frac{t}{2} (\gamma_s+3\gamma_t+2\gamma_c), \label{rge1}\\
\frac{d\gamma_t}{d\ln L} =&-\frac{t}{2} (\gamma_s-\gamma_t-2\gamma_c), \label{rge2}\\
\frac{d\gamma_c}{d\ln L} =&-\frac{t}{2} (\gamma_s-3\gamma_t) \label{rge3}.
\end{align}

\subsection{F.\, Relation to the BCS Hamiltonian}
\label{BCS}

Let us consider the interaction part of the Hamiltonian:
\begin{equation}
H_{\rm int} = \frac{1}{2} \int d\bm{r} d\bm{r^\prime} U(\bm{r}-\bm{r^\prime}) \bar{\psi}_\sigma(\bm{r}) {\psi}_\sigma(\bm{r})\bar{\psi}_{\sigma^\prime}(\bm{r^\prime})
{\psi}_{\sigma^\prime}(\bm{r^\prime})
\end{equation}
In the BCS case (for example, for short-range attraction mediated by phonons), 
$$U(\bm{r}-\bm{r^\prime}) = - \frac{\lambda}{\nu} \delta(\bm{r}-\bm{r^\prime}),$$ 
where the thermodynamic density of states $\nu$ {accounts for spin.}
According to Ref.~[S2]
the interaction parameters can be written as
\begin{equation}
\gamma_t =-\frac{F_t}{1+F_t}, \qquad \gamma_s=-\frac{F_s}{1+F_s},\qquad \gamma_c = -F_c 
\end{equation}
where
\begin{gather}
F_t = -\frac{\nu}{2} \langle U(2k_F\sin(\theta/2)) \rangle_{FS}, \qquad F_s = \nu U(q) + F_t, \qquad F_c = -\frac{\nu}{4} \langle U(2k_F\sin(\theta/2)) \rangle_{FS}-\frac{\nu}{4} \langle U(2k_F\cos(\theta/2)) \rangle_{FS} .
\end{gather}
Here $\langle\dots\rangle_{FS}$ denotes averaging over the Fermi surface. In the case of BSC Hamiltonian we find
\begin{equation}
F_t = \lambda/2, \qquad F_s = -\lambda/2,\qquad F_c = \lambda/2 
\end{equation}
and
\begin{equation}
\gamma_t \approx -\lambda/2, \qquad \gamma_s \approx \lambda/2,\qquad \gamma_c = -\lambda/2 .
\end{equation}

Thus, for the BCS case (i.e. when neither screened nor unscreened Coulomb repulsion is taken into account), 
we get the following interaction parameters at the ultraviolet scale (which is given by Debye frequency $\omega_D$ in the case of phonon-induced superconductivity): 
$$-\gamma_s=\gamma_t=\gamma_c = -\lambda/2.$$ 
As we will see in Sec. II below, this is precisely the relevant direction for the RG flow.
When disorder is weak, $\omega_D \tau \gg 1$, the initial values for these couplings in Eqs.~\eqref{rge1}-\eqref{rge3} are taken at the scale corresponding to the elastic scattering rate $1/\tau$. Then the Cooper interaction constant is renormalized at ballistic scales (between $1/\tau$ and $\omega_D$) such that
\begin{equation}
-\gamma_{s,0}=\gamma_{t,0}= -\lambda/2, \qquad \gamma_{c,0} = -\frac{\lambda/2}{1-(\lambda/2)\ln\omega_D\tau}=\frac{1}{\ln T_c^{BCS}\tau} 
\end{equation}
where  $T_c^{BCS} =\omega_D \exp(-2/\lambda)$.

The RG equations imply that there is the following perturbative correction to the BCS coupling due to interaction in the particle-hole channel
\begin{equation}
\gamma_c = \gamma_{c,0} -\frac{\lambda t_0}{2} \ln \frac{1}{T\tau},
\end{equation}
which is logarithmic in temperature. This leads to increase of  the transition temperature:
\begin{equation}
\frac{\delta T_c}{T_c^{BCS}} = \frac{t_0 \lambda}{2\gamma_{c,0}^2} \ln \frac{1}{T_c^{BCS}\tau} =\frac{t_0 \lambda}{2} \left (\ln \frac{1}{T_c^{BCS}\tau} \right )^3. \label{pertT}
\end{equation}
This perturbative calculation is justified in the regime $\delta T_c/T_c^{BCS} \ll 1$ which corresponds to the condition $t_0 \ll |\gamma_{c,0}|^3/\lambda$. 
In order to neglect the renormalization of $t$ due to weak localization disorder should be sufficiently weak, $t_0 \ll |\gamma_{c,0}$ (see Sec. II). Therefore, the perturbative correction to $T_c$ given by Eq.~\eqref{pertT} is valid for $t_0/|\gamma_{c,0}| \ll \min\{1, |\gamma_{c,0}|^2/\lambda\}$. 

In sufficiently dirty systems, $\omega_D \tau \ll 1$, the bare values of interactions constans in Eqs.~\eqref{rge1}-\eqref{rge3} are taken at the scale corresponding to $\omega_D$. In this case, 
\begin{equation}
\frac{\delta T_c}{T_c^{BCS}} = \frac{2t_0}{\lambda} \ln \frac{\omega_D}{T_c^{BCS}} = \frac{4 t_0}{\lambda^2} ,
\end{equation}
which holds for $t_0 \ll \gamma_{c,0}^2$. In the next section we will analyze the case of stronger disorder for which the perturbative treatment is insufficient and one should solve the full set of RG equations.

\section{II.\, Analysis of renormalization group equations}
\label{analysis}

We consider first 2D systems of the orthogonal symmetry class at weak
disorder ($g \gg 1$) where the localization and the multifractality
effects are weak. Then we turn to 2D systems of the symplectic symmetry class
in the metallic regime ($g\gg 1$) characterized by weak antilocalization. 
Finally, we address the vicinity of an Anderson
transition (symplectic class in 2D and orthogonal class in 3D).
In all these cases the multifractal character of wave functions in dirty
systems can strongly enhance the superconducting transition temperature as
compared to that of the clean system (usual BCS).

\subsection{A.\, 2D, orthogonal symmetry class}

\label{s2}

The full set of RG equations for the orthogonal symmetry class reads:
\begin{eqnarray}
\frac{d t}{d \ln L} &= & t^2 - \left(\frac{\gamma_s}{2} + 3
\frac{\gamma_t}{2} + \gamma_c\right)t^2, \label{e2.1}  \\[0.2cm]
\frac{d \gamma_s}{d \ln L} &=& - \frac{t}{2}
(\gamma_s + 3 \gamma_t + 2\gamma_c) , \label{e2.2}\\[0.2cm]
\frac{d \gamma_t}{d \ln L}& = & -\frac{t}{2} (\gamma_s - \gamma_t -
2\gamma_c) , \label{e2.3}\\[0.2cm]
\frac{d \gamma_c}{d \ln L}& = & -\frac{t}{2}(\gamma_s-3\gamma_t)-2\gamma_c^2, \label{e2.4}\\[0.2cm]
\frac{d \ln z}{d \ln L}&=&  \frac{t}{2} (\gamma_s + 3\gamma_t +
2\gamma_c)\,. \label{e2.5}
\end{eqnarray}
Here, we remind, $t\ll 1$ is the dimensionless resistance (inverse dimensionless
conductance,  $t=2/\pi g$), $\gamma_i  = \Gamma_i/z
 \ll 1$ are interaction constants
in the singlet (s), triplet (t) and Cooper (c) channels.
In Eq.~(\ref{e2.1}) the first term is due to weak localization. The
second term is due to interaction [Altshuler-Aronov (AA) correction], it
can be neglected within our accuracy. In
Eqs.~(\ref{e2.2})--(\ref{e2.4}) only terms linear in interaction
constants have been kept. The only exception is the Cooper term
$-2\gamma_c^2$ in Eqs.~(\ref{e2.4}) which is responsible for the BCS
superconducting transition. The neglected terms are much smaller
because they contain an additional small factor $t$. Below we shall use
$y= \ln L$ where $L$ is the running RG length scale. We measure lengths in units of the microscopic scale 
where the RG \eqref{e2.1}-\eqref{e2.5} starts.

Neglecting the AA contributions, we rewrite the set of equations
(\ref{e2.1})--(\ref{e2.4}) in the form
\bea
\label{e2.6}
\frac{d}{dy} \left( \begin{array}{c} \gamma_s \\ \gamma_t \\ \gamma_c
\end{array} \right) &=& -\frac{t}{2} \left( \begin{array}{ccc}
1 & 3 & 2 \\
1 & -1 & -2 \\
1 & -3 & 0\\
\end{array} \right)
\left( \begin{array}{c} \gamma_s \\ \gamma_t \\ \gamma_c
\end{array} \right) - 
\left( \begin{array}{c} 0 \\ 0 \\ 2\gamma_c^2
\end{array} \right) \,; \\[0.2cm]
\frac{dt}{dy} &=& t^2 \,.
\label{e2.7}
\eea
Eigenvalues of the matrix  in Eq.~(\ref{e2.6}) are $\lambda = -4$, $\lambda^\prime=2,
2$ (not including the prefactor $-t/2$); the corresponding
 eigenvectors are as follows:
\be
\label{e2.8}
\lambda=-4\ : \ \  \left( \begin{array}{c} -1 \\ 1 \\ 1
\end{array} \right)
\ ;\qquad \qquad
\lambda^\prime=2\ :
\left( \begin{array}{c} 1 \\ 1 \\ -1
\end{array} \right)
\ {\rm and} \
\left( \begin{array}{c} 1  \\ -1 \\ 2
\end{array} \right) .
\ee
If the $\gamma_c^2$ term is neglected, the solution of the linear
system (\ref{e2.6}) approaches the eigenvector with $\lambda=-4$,
i.e., $\gamma_s = - \gamma_t = - \gamma_c$.
Let us expand the vector formed by $\gamma_i$ in eigenvectors
(\ref{e2.8}):
\be
\label{e2.9}
\left( \begin{array}{c} \gamma_s \\ \gamma_t \\ \gamma_c
\end{array} \right) = a \left( \begin{array}{c} -1 \\ 1 \\ 1
\end{array} \right) +
b \left( \begin{array}{c} 1 \\ 1 \\ -1
\end{array} \right) +
c \left( \begin{array}{c} 1  \\ -1 \\ 2
\end{array} \right) =
\left( \begin{array}{ccc} -1 & 1 & 1 \\
                           1 & 1 & -1 \\
                           1 & -1 & 2
\end{array} \right)
\left( \begin{array}{c} a \\ b \\ c \end{array} \right) \,.
\ee
For convenience, we also present the inverted transformation:
\be
\label{e2.9a}
\left( \begin{array}{c} a \\ b \\ c
\end{array} \right) =
\left( \begin{array}{ccc} -{1/6} & {1/2} & {1/3} \\
                           {1/2} &{1/2}  & 0 \\
                           {1/3} & 0 & {1/3}
\end{array} \right)
\left( \begin{array}{c} \gamma_s \\ \gamma_t \\ \gamma_c
\end{array} \right) \,.
\ee

Transforming the set of equations (\ref{e2.6}) to the new variables
$a,b,c$, we get
\bea
\label{e2.10}
\frac{da}{dy} &=& 2 t a - \frac{2}{3} (a-b+2c)^2  ,\\
\frac{db}{dy} &=& - t b ,
\label{e2.11} \\
\frac{dc}{dy} &=& - t c - \frac{2}{3} (a-b+2c)^2 ,
\label{e2.12} \\
\frac{dt}{dy} &=& t^2 \,.
\label{e2.13}
\eea
Equations~\eqref{e2.10}-\eqref{e2.13} are supplemented by the following initial conditions: $a(0)=a_0$, $b(0)=b_0$, $c(0)=c_0$ and $t(0)=t_0$.
Equation (\ref{e2.13}) yield the standard evolution of the resistance
due to weak localization:
\be
\label{e2.14}
t(y) = \frac{t_0}{1 -t_0y} \,.
\ee
Equation  (\ref{e2.11}) can now be solved, yielding
\be
\label{e2.15}
b(y) = b_0(1 -t_0y) \equiv b_0 \frac{t_0}{t} \,.
\ee
Since $b$ decreases upon RG, it is not important and we neglect it in
the future analysis.

Equations for the remaining two variables, $a$ and $c$ are coupled. If
the quadratic term is neglected, then $a$ increases and $c$
decreases. This suggests that $c$ can be neglected. This is confirmed
by a more careful analysis which shows that, although on the very last
interval of RG ``time'' $y$ the variable $c$ starts to increase and
becomes of the same order as $a$ (i.e. of order unity), this 
weakly affect the RG scale at which this happens (i.e. the temperature of the
superconducting transition). Thus, we neglect $c$ in
what follows.

We can now easily solve the remaining equation for $a$. We assume the
starting value $a_0$ to be negative (which means that there is
attraction in the Cooper channel that is supposed to lead to the
superconductivity), $a = - |a|$. This is in particular the case when
$\gamma_{c,0}$ is the dominant coupling and $\gamma_{c,0}<0$. The
equation reads
\be
\label{e2.16}
\frac{d|a|}{dy}  = 2t|a| + \frac{2}{3} a^2 \,.
\ee 

Solving this equation, we obtain
\be
\label{e2.18}
a(y) = - \left( \frac{t_0^2}{|a_0|t^2} +  \frac{2 t_0}{3 t^2} 
- \frac{2}{3t}\right)^{-1} \, .
\ee

Let us analyze the obtained result. Let us first assume that
$|a_0| \ll t_0$. Then the second term in brackets in the r.h.s. of
(\ref{e2.18}) is small compared to the first one and can be neglected,
\be
\label{e2.19}
a^{-1}(y) = - \frac{1}{t} \left( \frac{t_0^2}{|a_0|t} -
  \frac{2}{3}\right) \,.
\ee
With increasing RG scale $y$ the 
resistance $t$ 
increases
together with the
interaction $a$. If $t$ reaches first unity, we get an
insulator; if $a \sim 1$ happens first, we get a superconductor. It is
easy to see that the second possibility (superconductivity) is
realized if  $|a_0| \gg t_0^2$.  Then at the point of the transition
to superconductivity we have a resistance
\be
\label{e2.20}
t_* \simeq \frac{3 t_0^2}{2 |a_0|} \ll 1 \,.
\ee
The transition will then happen at
\be
\label{e2.21}
y_* \simeq \frac{1}{t_0} - \frac{1}{t_*}
\ee
i.e. at temperature
 \be
\label{e2.22}
T_c^* \sim \exp\left\{- \frac{2}{t_0}\left [1-\frac{t_0}{t_*}\right ]
\right\} \, .
\ee
Here the factor 2 in the exponent originates from a translation 
of the length scale into energy (temperature).
Under the above assumption  $|a_0| \ll t_0$ the second term in square
brackets in the exponential of (\ref{e2.22}) is just a small correction to the first
one. By solving \eqref{e2.12} with $b=0$ and $a$ given by Eq.~\eqref{e2.19}, one finds that although $|c|$ decreases initially, eventually with increasing RG scale towards $y_*$ it becomes of the order of unity: $|c(y_*)|\sim 1$. Therefore, to determine precise value of $t_*$ one has to solve coupled equations for $a$ and $c$ (with $(b=0)$).

The transition temperature~\eqref{e2.22} is much
higher than the  BCS temperature $T_{\rm BCS} \sim
e^{-1/|\gamma_{c,0}|}$, so that the superconductivity is strongly enhanced by
disorder. The origin of the enhanced superconductivity is in the
increase of $|a|$
governed by the eigenvalue $\lambda =-4$ of Eq.~(\ref{e2.8}) which
yields the eigenvalue $-(t/2) \times (-4) = 2t$ of the linear
part of the system (\ref{e2.6}). This is nothing but the anomalous
multifractal exponent $-\Delta_2$ for this symmetry class. (We have in
mind the ``weak multifractality'' in 2D.) Therefore,
the (multi)fractality is the source of the enhancement of the
superocnductivity.

It is worth mentioning that there is no such enhancement in the absence of 
interaction in particle-hole singlet and triplet channels.
Indeed, the interplay of disorder and interaction in the renormalization
of $\gamma_c$ does not produce the term $t\gamma_c$ on the r.h.s. Eq.~(\ref{e2.4}).
Of course, if initially absent, the triplet and singlet amplitudes are generated
due to the interplay of the Cooper amplitude and disorder, see Eq.~(\ref{e2.2}) and 
(\ref{e2.3}). Furthermore, as we have seen in Sec. I\,F above, the BCS interaction
in fact contributes to all interaction channels, so that even in the absence 
of electron-electron repulsion the bare values of $\gamma_s$ and $\gamma_t$
are non-zero.

If $|a_0| \ll t_0^2$, the resistance reaches unity before the
interaction becomes strong, and the system is an insulator. Finally,
if $|a_0| \gg t_0$, the disorder is not particularly important, and
the transition temperature is given by usual clean BCS, $T_* \sim
e^{-1/|\gamma_{c,0}|}$. (In the latter case neglecting $b$ and $c$ is not
parametrically justified and leads to an incorrect numerical factor in
the exponent.)

\subsection{B.\, 2D, symplectic symmetry class, metallic regime}
\label{s3}

Now we consider the symplectic class, i.e. assume that the spin
symmetry is broken 
for example, by sufficiently strong spin-orbit
interaction. This leads to the following two modifications: (i) weak
antilocalization rather than weak localization, and (ii) triplet
interaction channel is suppressed and can be discarded. With these
modifications, the system  (\ref{e2.1})--(\ref{e2.4}) becomes (cf. Ref.~[S4])
\begin{eqnarray}
\frac{d t}{d y} &= & - \frac{1}{2} t^2 - \left(\frac{\gamma_s}{2} +
\gamma_c\right)t^2,
\label{e3.1}  \\[0.2cm]
\frac{d \gamma_s}{d y} &=& - \frac{t}{2}
(\gamma_s + 2\gamma_c) ,
\label{e3.2} \\[0.2cm]
\frac{d \gamma_c}{d y}& = & -\frac{t}{2}\gamma_s-2\gamma_c^2 \,.
\label{e3.3}
\end{eqnarray}
Similar equations for the symplectic case of long-range Coulomb interaction have been derived 
in Ref.~[S5].
The Altshuler-Aronov terms in (\ref{e3.1}) can again be
  neglected. Equation (\ref{e3.1}) then yields the standard
  antilocalization behavior,
\be
\label{e3.4}
t(y) = \frac{t_0}{1+ yt_0/2} \,.
\ee
The system of equations (\ref{e3.1}), (\ref{e3.2}) in the matrix form
is:
\be
\label{e3.5}
\frac{d}{dy} \left( \begin{array}{c} \gamma_s \\  \gamma_c
\end{array} \right) =
 -\frac{t}{2} \left( \begin{array}{ccc}
1 &  2 \\
1 & 0\\
\end{array} \right)
\left( \begin{array}{c} \gamma_s \\ \gamma_c
\end{array} \right) -
\left( \begin{array}{c} 0 \\ 2\gamma_c^2
\end{array} \right) \,.
\ee
Eigenvalues of the matrix  in Eq.~(\ref{e3.5}) are $\lambda = -1$, $\lambda^\prime=
2$ (not including the prefactor $-t/2$); the corresponding
 eigenvectors are as follows:
\be
\label{e3.6}
\lambda=-1\, : \quad  \left( \begin{array}{c} -1 \\ 1
\end{array} \right)
\ ;\qquad \qquad
\lambda^\prime=2\, :
\quad \left( \begin{array}{c} 2 \\ 1
\end{array} \right) \,.
\ee
If the $\gamma_c^2$ term is neglected, the solution of the linear
system  approaches the eigenvector with $\lambda=-1$,
i.e., $\gamma_s = - \gamma_c$.
As in the orthogonal case, we can expand the vector formed by
$\gamma_i$ in eigenvectors
\be
\label{e3.7}
\left( \begin{array}{c} \gamma_s \\ \gamma_c
\end{array} \right) = a \left( \begin{array}{c} -1 \\ 1
\end{array} \right) +
c \left( \begin{array}{c} 2 \\ 1
\end{array} \right)  =
\left( \begin{array}{cc} -1 & 2 \\
                          1 & 1
\end{array} \right)
\left( \begin{array}{c} a \\ c \end{array} \right) \,.
\ee
 The inverted transformation is given as:
\be
\label{e3.8}
\left( \begin{array}{c} a \\ c
\end{array} \right) =
\left( \begin{array}{cc} - {1/3} &  {2/3}  \\
                           {1/3} &   {1/3}
\end{array} \right)
\left( \begin{array}{c} \gamma_s \\ \gamma_c
\end{array} \right) \,.
\ee

Transforming the set of equations \eqref{e3.7} to the new variables a, c, we find
\begin{align} \label{e3.8a}
\frac{da}{dy}  =& \frac{t}{2} a - \frac{4}{ 3} (a+c)^2, \\
\frac{dc}{dy}  =& - t c - \frac{2}{ 3} (a+c)^2 .
\label{e3.8b}
 \end{align}
Equations for two variables, $a$ and $c$ are coupled. If the quadratic term is neglected, then $a$ increases 
and $c$ decreases. At the later stage of RG the quadratic terms leads to enhancement of $c$. 
This suggests that $c$ can be neglected for qualitative analysis of RG equations~\eqref{e3.8a}-\eqref{e3.8b}. 
We thus neglect $c$ and keep only $a$ (fully analogously to what we have done in the
orthogonal case). The resuting equation for $a$ reads
\be
\label{e3.9}
\frac{d|a|}{dy}  = \frac{t}{ 2} |a| + \frac{4}{ 3} a^2 \,.
\ee
We solve this equation with the
result
\be
\label{e3.10}
a = \frac{1}{t} \left ( \frac{1}{ a_0 t_0} + \frac{4}{ 3 t_0^2} - \frac{4}{3 t^2}\right )^{-1} \,.
\ee
The new (different from usual BCS) behavior emerges under the
condition $a_0 \ll t_0$. Then the condition $a \sim 1$ yields
\be
\label{e3.11}
t_* \simeq 2 \left(a_0 t_0/3\right)^{1/2} \ll t_0 \,.
\ee
By solving Eq.~\eqref{e3.8b} with $a$ given by Eq.~\eqref{e3.10}, we 
find that although $|c|$ decreases initially, eventually with increasing RG scale it reaches $a$: $c\sim a$ at $t=t_*$. Therefore, to determine precise value of $t_*$ one has to solve coupled equations for $a$ and $c$.

Equation~\eqref{e3.11} yields the transition temperature 
\be
\label{e3.12}
T_c^* \sim e^{-2y_*}\sim e^{-4/t_*} \sim \exp\left\{ - \mathcal{C}/\left(a_0 t_0\right)^{1/2} \right\} \,,
\ee
much higher than $T_{\rm BCS} \sim e^{-1/|\gamma_{c,0}|}$. The BCS
behavior is restored (up to corrections) at  $a_0 \gg t_0$. The constant $\mathcal{C}$ is of the order of unity and depend on ration $c_0/a_0$.

As in the orthogonal case, the source of the enhancement of the
superconducting temperature is in the first term on the r.h.s. of
eq.~(\ref{e3.9}). The eigenvalue $t/2$ is  the anomalous
multifractal exponent $-\Delta_2$ for the symplectic symmetry
class. Therefore, also in this case
the (multi)fractality is the source of the enhancement of the
superocnductivity. This enhancement is less efficient than in the
orthogonal case for two reasons, because of antilocalizing behavior
that leads to decrease of $t$ and therefore weakening of
multifractality.

\subsection{C.\, System at  or near Anderson transition}
\label{s4}

\subsubsection{1.\, Exactly at criticality}
\label{s4.1}

We now consider a system at the Anderson transition point. This may be
2D or 3D symplectic class system, or 3D orthogonal class. In all the
cases after the first (fast) part of the RG evolution, where the
Cooper term is assumed to be not important yet, the coupling constant
$\gamma_s$ and (in the orthogonal case) $\gamma_t$
``adjust'' to $\gamma_c$ according to the $\gamma_s = - \gamma_t = - \gamma_c$
(orthogonal) or  $\gamma_s = - \gamma_c$ (symplectic). So, the
main part of evolution can be described by a single equation, in
analogy with Eqs.~(\ref {e2.16}) and (\ref{e3.9}):
\be
\label{e4.1}
\frac{d\gamma}{ dy} = -\Delta_2 \gamma - a\gamma^2 \, , \quad a\sim 1
\ee

The superconductivity will take place if $\gamma_{0} < 0$. We also
note that $\Delta_2 < 0$. In particular, $\Delta_2= - 1.7 \pm 0.05$ for the 3D
orthogonal-class [S6]
 and $\Delta_2= - 0.344\pm 0.004$ for the 2D symplectic-class Anderson transitions [S7].
If  $|\gamma_{0}| \ll 1$, the second term in the r.h.s. of
(\ref{e4.1}) is not important for the evaluation of the leading
behavior of the transition temperature. Keeping only the first term,
we find
\be
\label{e4.2}
T_c^* \sim \exp\{-d y_*\} \sim |\gamma_{0}|^{d/|\Delta_2|} \,.
\ee
The factor $d$ in the exponential, where $d$ stands for the spatial dimensionality, translates length in temperature.
We see that at the Anderson transition point the enhancement of the
superconducting transition temperature becomes even stronger than in
2D: the transition temperature is now a power-law (rather than
exponential) function of the coupling constant $\gamma_{0}$.
Equation (\ref{e4.2}) agrees with the result of Ref.~[S3].

\subsubsection{2.\, Slightly off criticality: Insulating side}
\label{s4.2}

In the above consideration we assumed that the system is exactly at
the Anderson transition point. Let us analyze what happens if the
system is slightly off criticality. Then, we need to add the following equation to Eq.~\eqref{e4.1}:
\begin{equation}
\frac{d t}{dy} =\frac{1}{\nu}(t-t_c) + \eta \gamma . \label{e4.2a}
\end{equation}
Here, we take in account that the presence of interaction drives the system away from the non-interacting critical point. We note that the correlation length exponent $\nu>0$.  In particular, $\nu= 1.57\pm 0.02$ for the 3D
orthogonal-class [S8]
and $\nu=2.746 \pm 0.009$ for the 2D symplectic-class Anderson transitions [S9].

Neglecting the term of the second order in $\gamma$, the system of Eqs.~\eqref{e4.1} and \eqref{e4.2a} can be solved for $|\Delta_2|\nu \neq 1$ as
\begin{equation}
t = \tilde t + \frac{\eta \nu}{|\Delta_2|\nu-1} \gamma, \qquad  
\tilde t =t_c + \Bigl ( \tilde t_0-t_c\Bigr )e^{y/\nu} , \quad 
\gamma=\gamma_0 e^{|\Delta_2| y}, \qquad \tilde t_0 = t_0 -  \frac{\eta \nu\gamma_0}{|\Delta_2|\nu-1} .
\end{equation} 
In the special case $|\Delta_2|\nu=1$, we find
\begin{equation}
\tilde t = t - \eta \gamma_0 \gamma y, \quad \tilde t_0 = t_0.
\end{equation}
Therefore, the presence of the term $\eta \gamma$ in Eq.~\eqref{e4.2a} indicates that the proper scaling variable is $\tilde t$ rather than $t$. In what follows we shall omit `tilde' sign.

Up to the scale of the localization length
\be
\label{e4.3}
\xi \sim |t_0 - t_c|^{-\nu}
\ee
the RG will proceed as at the critical point. So, there are two
possibilities. If $L_* < \xi$, where $L_* \sim
|\gamma_{0}|^{-1/|\Delta_2|}$ is the length scale where the
superconducting transition at criticality takes place
(Sec.~II\,C\,1), then the transition temperature is not affected by
detuning. In the opposite case, the localization takes place first,
and there is no superconductivity. Thus, the condition ($\delta_\xi \propto \xi^{-d}$)
\be
\label{e4.4}
\delta_\xi \sim T_c^* \,,
\ee
where $T_c^*$ is given by Eq.~(\ref{e4.2}), is the condition of the
superconductor-insulator transition (at zero temperature). It is worth noting that Ref.~[S3]
argues that supeconducting state with $T_c \ll T_c^*$ persists further in the localized regime  ($\delta_\xi \gg T_c^*$) due to Mott-type rare configurations. Our RG approach (at least in its present form) is not sufficient to explore this posibility.

\subsubsection{3.\,Slightly off criticality: Metallic side}
\label{s4.3}

As in Sec.~II\,C\,2, we have the length scale $\xi$ given by
Eq.~(\ref{e4.3}) but it now has a meaning of the correlation
length. Below this scale the system is at criticality, above this
scale it becomes a metal. As on the localized side, if $\delta_\xi \ll
T_*$, the detuning from criticality is immaterial, and the transition
temperature is given by $T_*$, Eq.~(\ref{e4.2}). In the opposite case,
the result depend on whether we are in 2D symplectic case or in 3D
(4D, ...) orthogonal class.

\subsubsection{4.\, 3D, 4D, $\dots$, orthogonal class} 
\label{s4.3.1}

After the first step of RG (up to the scale $\xi$) the Cooper-channel
interaction constant takes the value
\be
\label{e4.5}
\tilde{\gamma}_{0} = \gamma_{0} \xi^{|\Delta_2|} \,.
\ee
After this the RG proceeds according to the usual BCS.
The total RG scale $y$ where the coupling becomes of order unity (and
thus the transition happens) is
\begin{equation}
\label{e4.6}
y_*^{(\xi)} = \ln\xi+\frac{c}{|\gamma_0|\xi^{|\Delta_2|}} ,
\end{equation}
where $c \sim 1$.Thus, the transition temperature is
\be
\label{e4.7}
T_*^{(\xi)} \sim \frac{1 }{ \xi^d} \exp\left\{-\frac{cd }{
    |\gamma_{0}|\xi^{|\Delta_2|}}\right\}  =
\delta_\xi \exp\left\{- cd \left(\frac{\delta_\xi }{
    T_*}\right)^{|\Delta_2|/d} \right\} \,.
\ee
It is easy to see that this intermediate regime correctly matches
results for two regimes between which it is located: at $\delta_\xi
\sim T_*$  we have $T_*^{(\xi)} \sim T_*$, and at $\delta_\xi \sim 1$
we have $T_*^{(\xi)} \sim T_{\rm BCS}$.

\subsubsection{5.\, 2D, symplectic class}
\label{s4.3.2}

After the first (Anderson-critical) step of RG the interaction
constant is given by the same formula (\ref{e4.5}) as in 3D. The
difference is on the second step: now we have to apply the formula for
a 2D symplectcic metal (\ref{e3.12}) with $t_0 \sim 1$. This yields
\be
\label{e4.8}
T_*^{(\xi)} \sim \frac{1}{ \xi^2} \exp\left\{-\frac{\mathcal{C}}{
    |\gamma_{0}|^{1/2}\xi^{|\Delta_2|/2}}\right\}  =
\delta_\xi \exp\left\{- \mathcal{C} \left(\frac{\delta_\xi }{
    T_*}\right)^{|\Delta_2|/4} \right\} \,.
\ee

\begin{itemize}

\item[]
\item[] 
\item[] 

\item[[S1\!\!]] 
A.M. Finkel'stein. {\it Electron Liquid in Disordered Conductors}, vol. 14 of Soviet Scientific Reviews, ed. by I.M.\,
Khalatnikov, Harwood Academic Publishers, London, (1990). 
\item[[S2\!\!]] 
D. Belitz and T.R. Kirkpatrick, {\it The Anderson-Mott transition}, Rev. Mod. Phys. {\bf 66}, 261 (1994).
\item[[S3\!\!]] 
M.V.~Feigelman, L.B.~Ioffe, V.E.~Kravtsov, and
  E.~Cuevas, Ann. Phys. {\bf 325}, 1390 (2010). 
\item[[S4\!\!]]
P. M. Ostrovsky, I. V. Gornyi, and A. D. Mirlin,
Phys. Rev. Lett. {\bf 105}, 036803 (2010). 
\item[[S5\!\!]] 
C.\
Castellani, C.\ DiCastro, G.\ Forgacs, and S.\ Sorella, Solid
State Comm. \textbf{52}, 261 (1984); M.\ Ma and E.\ Fradkin, Phys.
Rev. Lett. \textbf{56}, 1416 (1986). 
\item[[S6\!\!]] 
A. Mildenberger, F. Evers, and A. D. Mirlin, Phys.
Rev. {\bf B} 66, 033109 (2002). 
\item[[S7\!\!]] 
A. Mildenberger and F. Evers, Phys. Rev. B {\bf 75},
041303(R) (2007). 
  \item[[S8\!\!]] 
K. Slevin and T. Ohtsuki, 
Phys. Rev. Lett. \textbf{82}, 382 (1999).  
\item[[S9\!\!]] 
Y. Asada, K. Slevin, and T. Ohtsuki, Phys. Rev. Lett.
{\bf 89}, 256601 (2002); Phys. Rev. B {\bf 70},
035115  (2004). 
\end{itemize}

\end{widetext}


\begin{thebibliography}{99}
\bibitem{FootnoteFeigelman} Ref.~\protect\cite{Feigelman07} argues that supeconducting state with $T_c\! \ll\! T_c^*$ persists further in the localized regime  ($\delta_\xi\! \gg\! T_c^*$) due to Mott-type rare configurations. Our RG approach (at least in present form) is insufficient to explore this possibility.
\end{thebibliography}
\end{document}